%% file: main.tex
\documentclass[runningheads]{llncs}
\usepackage[T1]{fontenc}
\usepackage{graphicx}

\input{Commands}  

\begin{document}

\title{PFD or PDF: Rethinking the Probability of Failure in Mitigation Safety Functions\\}
\titlerunning{PFD or PDF: ...}
\author{Hamid Jahanian}
\institute{FS Expert (TÜV Rheinland) \#266/16-SIS\\UGL, Sydney, Australia\\ 
\email{hamid.jahanian@ugllimited.com}\\
\url{linkedin.com/in/hamid-jahanian}}

\maketitle

\begin{abstract}
SIL (Safety Integrity Level) allocation plays a crucial role in defining the design requirements for Safety Functions (SFs) within high-risk industries. SIL is typically determined based on the estimated Probability of Failure on Demand (PFD), which must remain within permissible limits to manage risk effectively. Extensive research has been conducted on determining target PFD and SIL, with a stronger emphasis on preventive SFs than on mitigation SFs. In this paper, we address a rather conceptual issue: we argue that PFD is not an appropriate reliability measure for mitigation SFs to begin with, and we propose an alternative approach that leverages the Probability Density Function (PDF) and the expected degree of failure as key metrics. The principles underlying this approach are explained and supported by detailed mathematical formulations. Furthermore, the practical application of this new methodology is illustrated through case studies.

\end{abstract}

\section{Introduction}\label{Sec_Intro}

Probability of dangerous Failure on Demand (PFD) and frequency of dangerous failure (PFH) are two continuous measures used by functional safety standards, such as IEC 61508 \cite{Ref_183}, to quantify the likelihood of failure in Safety Functions (SFs). These measures are also linked to the Safety Integrity Level (SIL), which represents the significance of the SF by a discrete number ranging from 1 to 4.

SIL allocation studies are conducted to assess the risk and determine the appropriate PFD/PFH and SIL. Quite commonly, a SIL allocation begins with determining the target PFD/PFH, which are then used to derive the corresponding target SIL based on the mapping tables given in the standard \cite{Ref_184}. By determining the ``right'' targets, SIL allocation studies help avoid unnecessary implementation costs and compromised safety.

\begin{figure}[!h]
\begin{center} \includegraphics[scale=0.5]{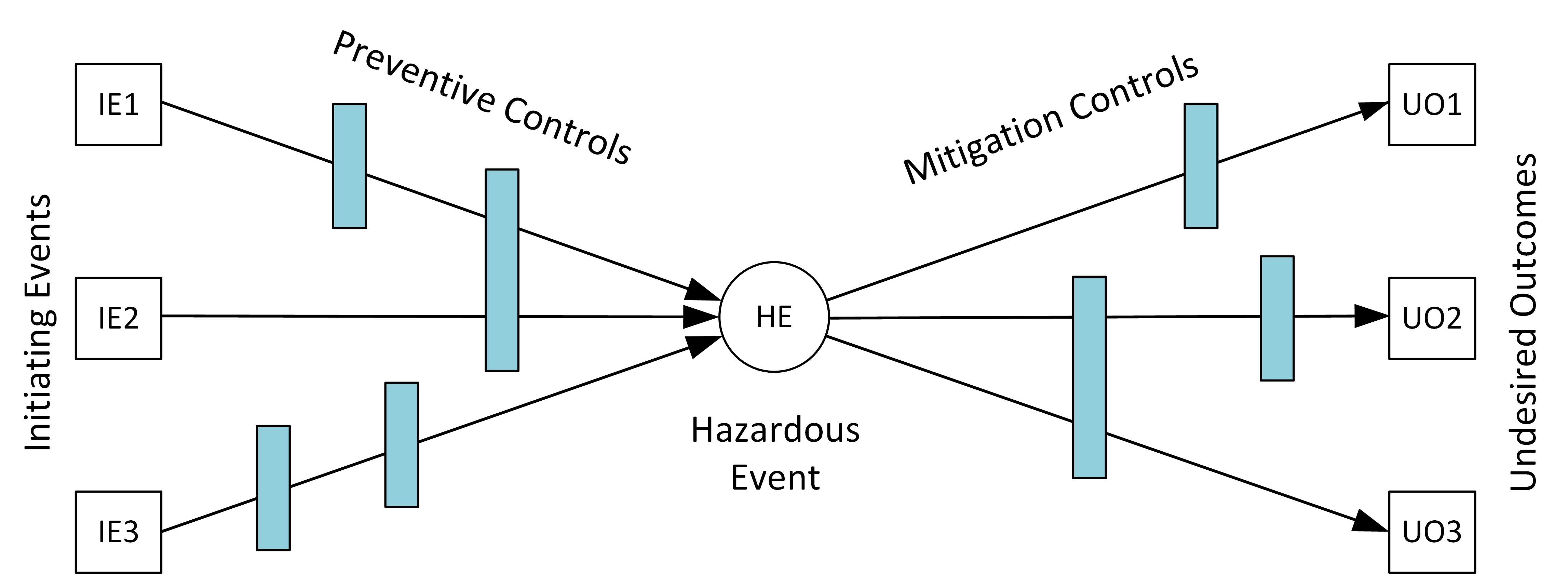} \end{center} \caption{Bowtie Accident Model} \label{Fig_Bowtie} \end{figure}

Safety Functions are classified into two primary categories: preventive functions and mitigation functions. Preventive functions operate before a Hazardous Event (HE) occurs, aiming to prevent it from happening. Conversely, mitigation functions act after an HE has occurred, focusing on minimising its consequences. \Fig{Fig_Bowtie} depicts a typical accident scenario in the form of a Bowtie diagram, with initiating events and preventive functions on the left, the hazardous event in the centre, and mitigation functions and undesired outcomes on the right.

With respect to SIL allocation, the safety standards apply the same concept of probability (PFD, PFH) to both preventive and mitigation functions, while recognising that different methods may be needed depending on the type of function \cite{Ref_187}. In this paper, we argue that the concept of probability of failure is not suitable for mitigation SFs, and proper analysis of these functions demands a different approach at both the conceptual and methodological levels. 

This need stems largely from the way in which preventive and mitigation functions respond to risk. Risk has two components: likelihood and severity of consequence. Preventive SFs help reduce the likelihood component, while mitigation functions help reduce the severity level. The performance of mitigation functions can be measured by the degree to which they \emph{reduce} the impact, which is inherently a non-binary performance. However, the success of preventive functions is judged by their binary role in \emph{preventing} an event, irrespective of its subsequent outcomes. We show that such inherent differences make the concept of PFD -- and thus SIL -- inaccurate for mitigation SFs, and we propose a new approach based on Probability Density Function (PDF) and expected degree of failure to more accurately measure the reliability of these functions.

The structure of our paper is as follows: \Sec{Sec_Background} defines the problem that this paper attempts to address. \Sec{Sec_MonBinFunc} begins the formal discussion and mathematical formulation by separating binary and non-binary functions. The formulation continues in \Sec{Sec_ExpRisk} by covering the concept of expected risk, and in \Sec{Sec_RiskMitig} by examining the impact of mitigation functions on risk. In \Sec{Sec_MitigTargets}, we conclude the formulation by proposing a procedure for setting reliability targets. This is then supplemented by some final thoughts in \Sec{Sec_FinalPoints}. To demonstrate the practical application of our method, \Sec{Sec_CaseStudy} presents two case studies -- one from a flammable gas ventilation system and another from a road tunnel fire scenario. \Sec{Sec_Discussion} explores additional aspects of SIL in relation to our proposed approach and outlines directions for future research. Finally, \Sec{Sec_Conclusion} summarises and concludes our contribution.

The following notations are used in this paper:
\begin{align}
l &: \textit{ number of subsystems} \nonumber \\
m &: \textit{ number of system functions} \nonumber \\
n &: \textit{ number of consequence segments} \nonumber \\
\Bold{F}=[f_{jk}]_{l \times m} &: \textit{ mapping subsystems to functions} \nonumber \\
\Bold{w}=[w_{h}]_{1 \times n} &: \textit{ estimated consequence frequencies} \nonumber \\
\Bold{\overline{w}}=[\overline{w}_{h}]_{1 \times n} &: \textit{ tolerable consequence frequencies} \nonumber \\
\Bold{c}=[c_{h}]_{1 \times n} &: \textit{ consequence severity values} \nonumber \\
r &: \textit{ collective estimated level of risk} \nonumber \\
\overline{r} &: \textit{ collective tolerable level of risk} \nonumber \\
w_{IE} &: \textit{ frequency of the initiating event} \nonumber \\
w_{HE} &: \textit{ frequency of the hazardous event} \nonumber \\
c_{HE} &: \textit{ hazardous event consequence} \nonumber \\
\Bold{q}=[q_j]_{1 \times l} &: \textit{ degrees of subsystems' failure} \nonumber \\
\Bold{s}=[s_k]_{1 \times m} &: \textit{ degrees of functions' success} \nonumber \\
\Bold{u}=[u_k]_{1 \times m} &: \textit{ contribution values of functions} \nonumber \\
\RV{X} &: \textit{ a random variable} \nonumber \\
E[\RV{X}] &: \textit{ expected value of } \RV{X} \nonumber \\
\overline{E[\RV{X}]} &: \textit{ tolerable level of the expected value of } \RV{X} \nonumber \\
f_{\RV{X}}(x) &: \textit{ Probability Density Function of } \RV{X} \nonumber
\end{align}

\section{The problem}\label{Sec_Background}

IEC 61508 sets a general framework for functional safety. The terms SIL and Safety Function are two core elements of this framework. As per the standard, SIL is a property of a Safety Function, and every Safety Function should be allocated a SIL. The standard considers mitigation functions as one category of SFs (clause 3.4.1 of \cite{Ref_186}):

\begin{quote}
``\textit{A safety-related system may ... be designed to mitigate the effects of the harmful event, thereby reducing the risk by reducing the consequences};''
\end{quote}

Therefore, a mitigation SF needs to be allocated a target SIL. In its Part 5, the standard even directly addresses the allocation of SIL for mitigation functions (clause A.9 in \cite{Ref_187}):

\begin{quote}
``\textit{When determining the safety integrity requirements it should be recognised that when making judgements on the severity of the consequence, only the incremental consequences should be considered. That is, determine the increase in the severity of the consequence if the function did not operate over that when it does operate as intended. This can be done by first considering the consequences if the system fails to operate and then considering what difference will be made if the mitigation function operates correctly. In considering the consequences if the system fails to operate there will normally be a number of outcomes all with different probabilities. Event tree analysis (ETA) may be a useful tool for this.}''
\end{quote}

The only issue, however, is that the standard does not provide details on how this can be done in practice. Conventional methods, such as Layers of Protection Analysis (LOPA) \cite{Ref_352}, Fault Tree Analysis (FTA) \cite{Ref_168}, and Event Tree Analysis (ETA) \cite{Ref_204}, each have their own restrictions and/or downsides. Jahanian covered these limitations in detail in \cite{Ref_382}. The paper also proposed a new method based on inclusive system state analysis, which overcomes fundamental restrictions, such as the assumption of independence between protection layers and the aggregation of multiple consequence segments.

Nonetheless, the problem of SIL allocation extends beyond the methodological aspect. There are fundamental questions that need to be addressed. The current definitions that support the concepts of PFD and SIL are based on the prevention of an event, which is clearly different from mitigating the impacts of an event. It is therefore worthwhile to revisit these concepts and examine their suitability for mitigation functions. This is what this paper aims to do.

Preventive and mitigation SFs differ in several ways. However, one particular distinction has a significant conceptual impact: the way in which the success and failure of functions are characterised. Preventive functions are binary in nature -- their success and failure can be described by a single binary state: 0 for failure, 1 for success. In contrast, the success and failure of mitigation functions are proportional, as they aim to \emph{reduce} the severity of consequences \emph{as much as} possible. 

The challenge is that a non-binary random variable (e.g., the performance of a mitigation function) cannot be represented by a single probability value (e.g., PFD). Allocating PFD to a function requires a binary distinction between its success and failure. The question then is whether such a distinction is useful, or even possible, in mitigation functions. 

More specifically, the question we attempt to address in this paper is the following: can a single probability value (PFD) effectively represent the probabilistic performance of a mitigation function; and, if not, what is an alternative approach for measuring the reliability of such functions? This is a critical question. If PFD is not suitable for mitigation functions, then SIL will also be rendered irrelevant, because it is directly linked to PFD. Consequently, if SIL cannot be properly defined, it becomes difficult to determine the necessary requirements for compliance with the safety standard.

This paper examines the issue in detail and proposes a new approach to address it. Our method adopts the same risk tolerance concept used in current SIL allocation practices -- in which the gap between tolerable and actual risk levels determines the target reliability for the safety function. However, in our approach, these targets are based on PDF and the expected degree of failure, rather than solely on the probability of failure; and the targets are translated into practical design requirements, rather than merely probability values. The proposed solution is mathematically formulated, its application is demonstrated through case studies, and its advantages and limitations are discussed.

\section{Non-binary success}\label{Sec_MonBinFunc}

Not all systems and functions can have their success (or failure) expressed in binary terms. Consider, for instance, a smoke extraction damper comprising ten independent modules, all of which are supposed to open in a fire situation. A partial opening of the damper -- say, four out of ten modules -- is neither a complete success nor a complete failure. Instead, it represents a partial success, as it still contributes to the extraction of smoke, thereby mitigating the consequences, albeit not fully. 

In this paper, a \textbf{non-binary system} (or \textbf{function}) is one whose success/failure cannot be characterised in binary terms but rather as a \emph{degree} of success/failure. Conversely, a system (or function) whose success/failure is expressed in discrete binary terms is referred to as a \textbf{binary system} (or \textbf{function}).\footnote{We use \textbf{bold} font when introducing key terms.}

A simple example of a binary system is a light bulb, which either works or does not work. However, many real-world systems are more complex. Consider a water pump, for instance: the pump may completely stop delivering flow if its electric motor fails, but it may also experience reduced flow due to partial mechanical faults in its impeller. Here, the motor behaves as a binary subsystem, the impeller as a non-binary subsystem, and the overall pump system is non-binary because it can be affected by either subsystem. 

In general, every system may be characterised as non-binary, but not every system may fit in the binary category. It would not be incorrect to describe these systems as non-binary. In that respect, the difference between a motor and a ten-module damper is that the motor transitions from complete success to complete failure in a very short period, whereas the damper takes longer due to its many redundant, independent components working in parallel. Hence, the term ``non-binary'' is inclusive, whereas ``binary'' is not. 

Binary systems are easier to analyse since their state is either 0 (failure) or 1 (success). Non-binary systems, however, require more complex analysis, as their success level varies between 0 (complete failure) and 1 (complete success). In some analyses, non-binary systems are simplified into binary ones, solely to make calculations more manageable.

From a mathematical perspective, binary systems can be seen as a subset of non-binary systems; the discrete values 0 and 1 are part of the continuous range [0,1]. Therefore, a computation framework designed for non-binary calculations can handle both binary and non-binary cases. However, a framework limited to binary calculations cannot effectively model non-binary systems.

\begin{figure}[!ht] 
\centering 
\subfloat[Binary System\label{Fig_Prob1}]{\includegraphics[width=0.45\columnwidth]{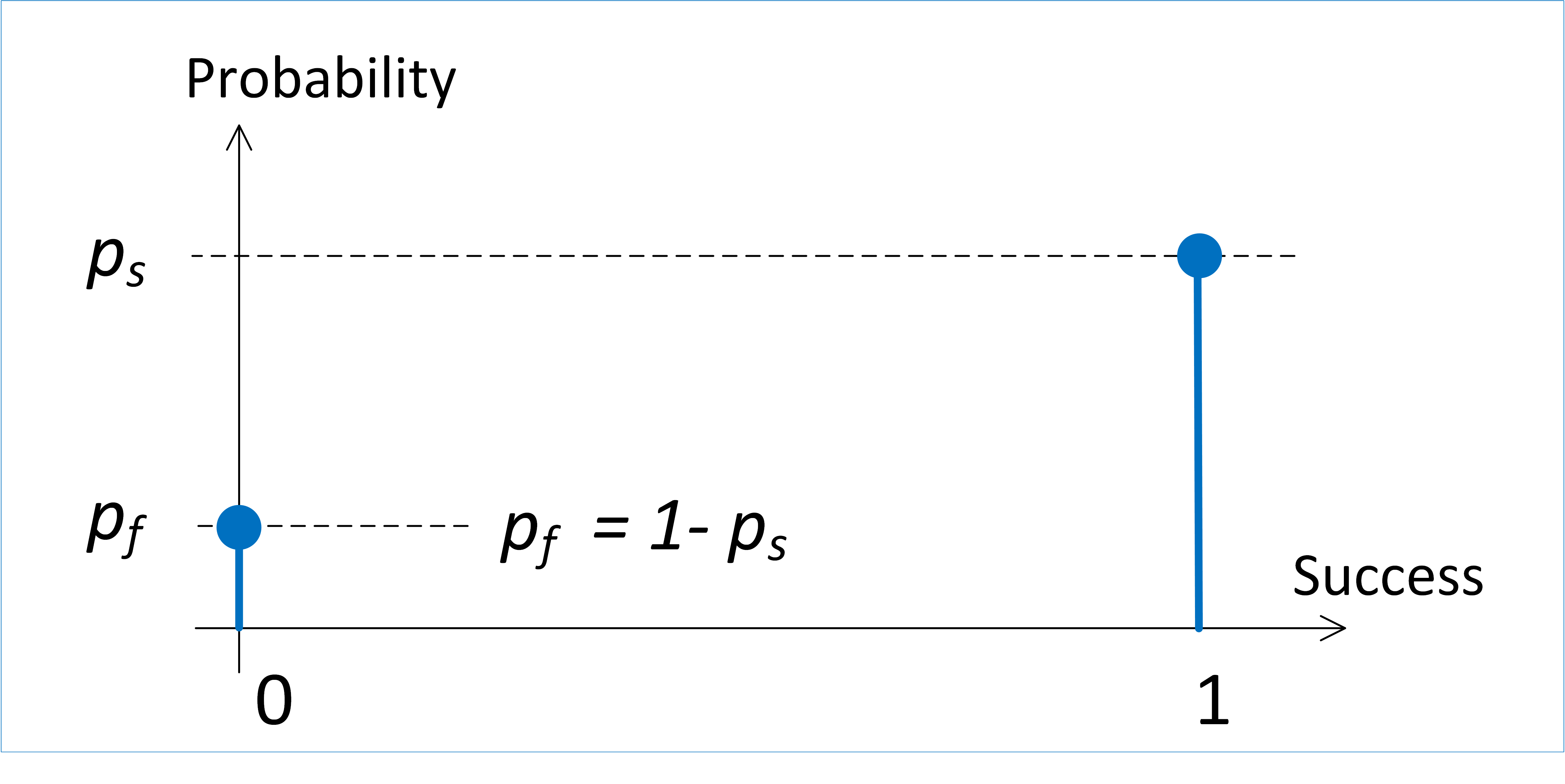}} 
\qquad 
\subfloat[Non-Binary System\label{Fig_Prob2}]{\includegraphics[width=0.45\columnwidth]{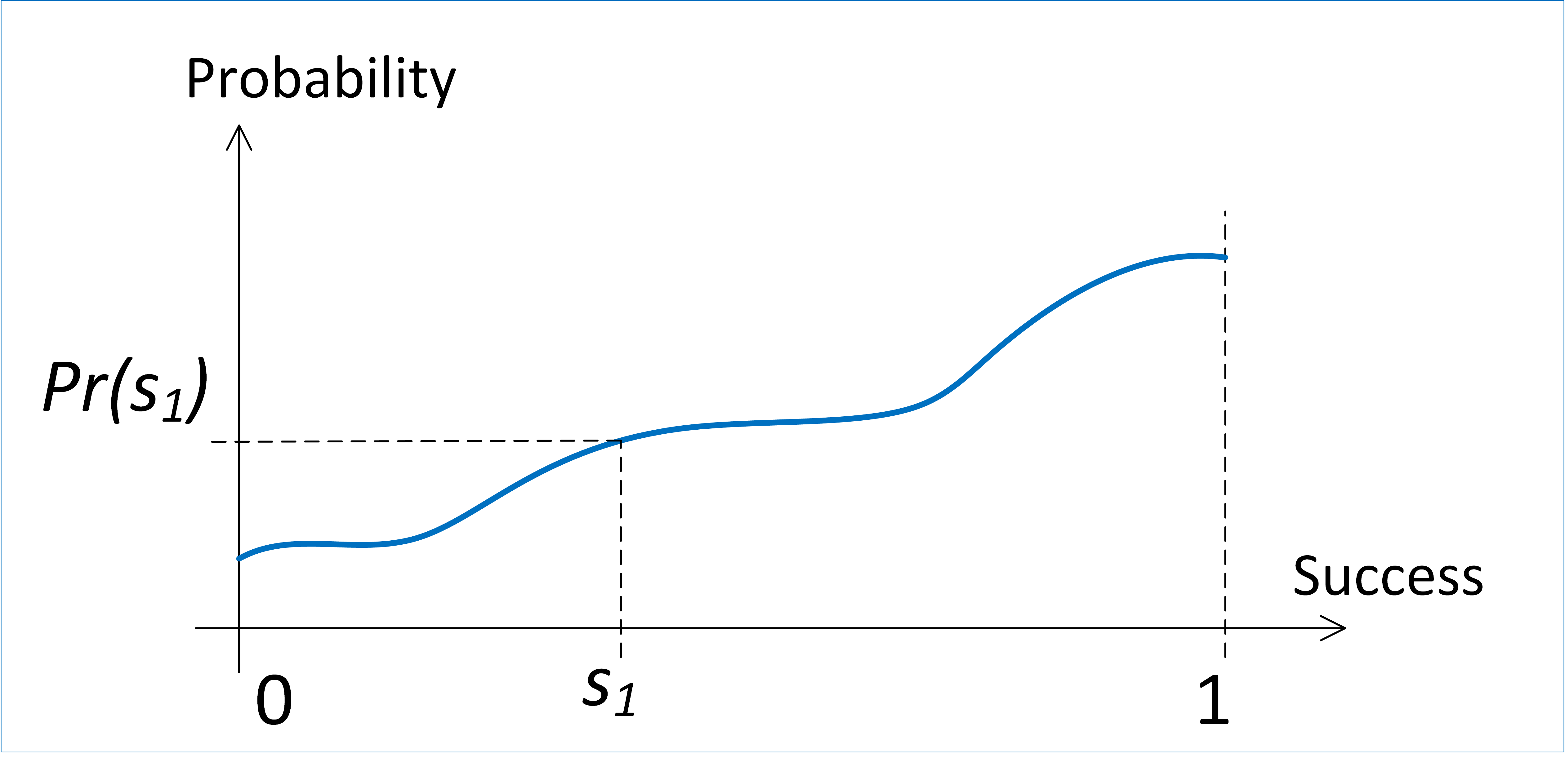}} 
\caption{Probability of System Success} \label{Fig_Prob} 
\end{figure}

Binary systems are particularly convenient for probability calculations. A single probability value suffices to describe the system’s state: if the probability of success is $p_s$, then the probability of failure is simply $p_f = 1 - p_s$, and no further information is required. In contrast, non-binary systems require the information in a complete Probability Distribution Function (PDF) to capture the range of all possible states between complete success and complete failure. The difference is illustrated in Figure~\ref{Fig_Prob}.  

Working with fully formulated PDFs in safety analyses such as SIL allocation can be challenging. However, this should not imply that the only alternative is to use an oversimplified probability value, namely the PFD. One way to simplify probability distribution in non-binary systems is to use the ``expected value'', which condenses the entire range of possible values of the random variable into a single representative value. For a continuous random variable $\RV{X}$, the expected value $E[\RV{X}]$ is calculated as:
\begin{align} &E[\RV{X}] =\int_{-\infty}^{\infty} f_{\RV{X}}(x) \cdot x \cdot dx \label{Eq_Expected} \end{align}
where $f_{\RV{X}}(x)$ is the PDF of $\RV{X}$. For a discrete random variable $\RV{X}$ with $n$ possible values, $E[\RV{X}]$ is calculated by:
\begin{align} &E[\RV{X}] =\sum_{i=1}^{n} Pr(x_i) \cdot x_i \label{Eq_Expected1} \end{align}
in which $Pr(x_i)$ is the probability of $x$ being $x_i$.

An expected value, also known as the ``first moment'', can be viewed as the centre of gravity of the corresponding random variable. It provides a concise measure that represents the overall behaviour of the random variable across its domain.

In reliability analysis, the success and failure of systems are random variables. We use $s \in [0, 1]$ and its corresponding random variable $\RV{S}$ to represent the \textbf{success} of a system (or function), and we use \Eq{Eq_Expected} or \Eq{Eq_Expected1} to calculate its expected value. Similarly, we use $q=1-s$ as the system (or function) \textbf{failure} measure, with $\RV{Q}$ as its corresponding random variable.\footnote{This is a general introduction of $s$ and $q$. In \Sec{Sec_RiskMitig}, $s$ is used exclusively for functions and $q$ is used exclusively for subsystems.}

Note that $E[\RV{S}]$ is not a probability value but rather the mean of $\RV{S}$. However, in the special case of binary systems, where failure and success are expressed as $s=0$ and $s=1$, respectively, the expected value of success equals the probability of success. Using \Eq{Eq_Expected1}:
\begin{align} 
&E[\RV{S}]=0 \times Pr(0) + 1 \times Pr(1) = Pr(1) \label{Eq_ExpProb}
\end{align}

Using expected values, both binary and non-binary systems can be analysed within the same computational framework. What differentiates them is not the failure measure but the way each transitions between states. In binary systems, the transition is a jump between the two states of success and failure. In non-binary systems, the transition is a continuous or multi-step travel within the range of complete success and complete failure.

It should also be noted that the success and failure discussed here refer solely to the effects of random faults, and are distinct from the ``coverage'' and ``effectiveness'' factors found in some sources \cite{Ref_368,Ref_362,Ref_363,Ref_364,Ref_365}.\footnote{Reader can refer to \cite{Ref_382} for a review summary of the cited sources.} Those characteristics refer to systematic limitations or external faults. For example, if a fire detector’s coverage drops off at five metres, it is not a random failure if the detector does not identify a fire ten metres away -- this simply reflects its design. Hence, success and failure in this paper always refer to random system faults, not systematic ones or external influences.

\section{Expected risk}\label{Sec_ExpRisk}

In industry, a preventive SF is a function that, if successful, prevents a Hazardous Event (HE) from occurring. Preventive SFs are typically binary; an event is either prevented or not prevented. In contrast, mitigation SFs generally fall into the category of non-binary functions, as they \emph{reduce} the consequences of the HE after it has occurred. Despite their different roles, preventive and mitigation functions share the same ultimate objective: to reduce risk. 

\textbf{Risk} is defined as the combination of likelihood and severity of consequences (see \cite{Ref_186}, for example). Likelihood may be expressed as either a probability or a frequency of occurrence, while severity can be expressed using any scale that quantifies the extent of damage (e.g., monetary value, injury scale, utility index). Where both likelihood and severity are expressed numerically, one common method of \emph{combining} them is multiplication. With $w_{HE}$ representing the frequency of occurrence of the HE, and $c_{HE}$ the consequence severity, the risk $r$ can be estimated as:
\begin{align}
&r = w_{HE} \cdot c_{HE} \label{Eq_RiskGeneral} 
\end{align}

When examining a preventive SF, the right-hand side of the Bowtie model (Figure~\ref{Fig_Bowtie}) can be condensed into a single value $c_{HE}$, since preventive SFs do not influence this part of the model. Conversely, when analysing mitigation functions, $w_{HE}$ is treated as a given, and the focus shifts to the variability of consequences on the right-hand side.

In mitigation analysis, the range of consequence severity may be divided into discrete parts, referred to as \textbf{segments}. For example, a severity range of 0 to 10 may be divided into five segments, each spanning 2 units. 

Let $n$ be the total number of consequence segments, and let $c_h$ and $w_h$ represent the consequence severity and estimated frequency of occurrence for segment $h$, respectively. The total risk can then be estimated as:
\begin{align}
&r=\sum_{h=1}^{n} w_h \cdot c_h = \sum_{h=1}^{n} w_{HE} \cdot Pr(c_h) \cdot c_h = w_{HE} \sum_{h=1}^{n} Pr(c_h) \cdot c_h \label{Eq_RiskDiscB}
\end{align}
Here, $Pr(c_h)$ is the probability that consequence severity $c$ equals $c_h$ if the HE occurs. Since only one severity level can manifest in any scenario, it also holds that $\sum_{h} Pr(c_h) = 1$.

Comparing the last sum in \Eq{Eq_RiskDiscB} to the sum in \Eq{Eq_Expected1}, we recognise it as the expected value of the consequence severity. Thus, \textbf{expected risk} will be:
\begin{align}
&r= w_{HE} \cdot E[\RV{C}] \label{Eq_RiskDiscB11}
\end{align}
where $\RV{C}$ is the random variable representing consequence severity, and $E[\RV{C}]$ is the \textbf{expected consequence severity}. 

One frequently raised question in risk analysis workshops is how to estimate the severity of consequences. Worst-case outcomes are usually too rare, while minor outcomes are more likely but less impactful. Relying solely on either extreme can distort the analysis, and using a midpoint value is often arbitrary. $E[\RV{C}]$ provides a better representation of the consequence profile by incorporating the full severity range via an expected value. The challenge then lies in estimating the $Pr(c_h)$ values, which may be derived from expert judgement in qualitative analyses or through statistical calculations in quantitative studies.

Another benefit of \Eq{Eq_RiskDiscB11} is that it applies equally to both discrete and continuous severity scales, as outlined in Section~\ref{Sec_MonBinFunc}. It is only the calculation of $E[\RV{C}]$ itself, that varies depending on whether the system is discrete or continuous, and for which either \Eq{Eq_Expected} or \Eq{Eq_Expected1} should be used.

When using \Eq{Eq_RiskDiscB11}, it is essential that all consequences are expressed on a common scale. For instance, if an accident leads to both a \$40M financial loss and two fatalities, these must be converted into a single severity scale -- such as a normalised utility scale from 0 to 1 -- so they can be aggregated.

~

Once the actual risk is estimated, it can be compared against a \textbf{tolerable risk} level $\overline{r}$. The objective is then to ensure that:
\begin{align}
&r \leq \overline{r} \label{Eq_TolRiskCrit}
\end{align}

Setting the tolerable risk criteria $\overline{r}$ is a compliance-driven process rather than a purely analytical one. It should reflect applicable legislation, national or organisational risk appetite, and other policy considerations. However, when tolerable frequencies are assigned to individual consequence segments, the overall tolerable risk can be qualified against:
\begin{align}
&\overline{r}=\sum_{h=1}^{n} \overline{w}_h  \cdot c_h  \label{Eq_RsikPDF_Tol}
\end{align}
where $\overline{w}_h$ is the tolerable frequency for segment $c_h$.

Given $\overline{r}$ and a known initiating frequency $w_{HE}$, we can determine the tolerable expected consequence severity $\overline{E[\RV{C}]}$:
\begin{align}
&\overline{E[\RV{C}]}=\overline{r} / w_{HE} \label{Eq_RiskDiscB111}
\end{align}
The requirement \Eq{Eq_TolRiskCrit} then becomes:
\begin{align}
&E[\RV{C}] \leq \overline{E[\RV{C}]} \label{Eq_RiskTolConseq}
\end{align}

It is important to note that tolerances defined for the expected value of consequence severity need not match the tolerances defined for each individual consequence segment. For example, if the tolerable frequency for a single fatality is $\V{1.0E-05}$ per year, then the tolerable risk for an expected consequence aggregating one fatality with multiple injuries could be higher. This is not because such outcomes are more tolerable, but because the measure now includes a sum of two scenarios with differing severity levels.

\section{Risk mitigation}\label{Sec_RiskMitig}

We use the term \textbf{mitigation function} to describe a function that reduces risk by decreasing the severity of consequences \cite{Ref_368}. A function that reduces the probability of occurrence of an accident is referred to as a \textbf{preventive function}. The term \textbf{mitigation system} refers to the complete system that enables the performance of a set of mitigation functions.

Consider a fire incident in a road tunnel, where fire and smoke may result in casualties and asset damage. Typical mitigation functions in this scenario include fire suppression, smoke extraction, and evacuation of road users. The mitigation system is the overall tunnel operation system that enables these functions to operate.

A mitigation system comprises a set of \textbf{subsystems}. Each mitigation function depends on one or more of these subsystems, and some subsystems may contribute to multiple functions. Subsystems are assumed to be independent. This assumption allows us to use the product rule in calculating system states and is fundamental to modelling function success.

Let the mitigation system consist of $l$ independent subsystems, supporting $m$ mitigation functions. We define the relationship between subsystems and functions using a \textbf{mapping matrix} $\Bold{F}$:
\begin{align}
&\Bold{F}=\bigl[f_{jk}]_{l \times m}, ~~ f_{jk} \in \{0, 1\} \label{Eq_MapF}
\end{align}
where $f_{jk}=1$ indicates that the $j$th subsystem is necessary for the $k$th function to succeed, and $f_{jk}=0$ indicates no dependency.\footnote{Some safety standards use simpler representations of safety functions and systems (e.g., IEC 61511 \cite{Ref_190}). However, these are primarily developed for preventive functions. In scenarios where multiple mitigation functions share subsystems, the use of a mapping matrix such as $\Bold{F}$ is essential to define interdependencies.}

Unlike preventive functions, the success or failure of mitigation functions is not strictly binary. Instead, it is proportional. We define $s_k$ as the \textbf{degree of success} of the $k$th function, ranging from 0 (complete failure) to 1 (complete success), and $\Bold{s}$ as the vector of all $s_k$s:
\begin{align}\label{Eq_FunctionS}
&\Bold{s}=[s_k]_{1 \times m}, ~~ 0 \leq s_k \leq 1
\end{align}

In most mitigation scenarios, multiple functions may be activated in parallel. We define $u_k$ as the \textbf{contribution} of the $k$th function to the overall mitigation effort -- that is, the proportion of risk reduction that depends solely on the $k$th function -- and $\Bold{u}$ as the vector of all $u_k$s:
\begin{align}
&\Bold{u}=[u_k]_{1 \times m}, ~~ 0 \leq u_k \leq 1, ~~ \sum_{k=1}^{m} u_k = 1
\end{align}

The contribution $u_k$ is a fixed (non-random) parameter indicating maximum functional importance. The actual success level $s_k$, however, is a random variable that depends on the success of the subsystems that perform function $k$.

We define $q_j$ as the \textbf{degree of failure} of subsystem $j$, and $\RV{Q}_j$ as its corresponding random variable:
\begin{align}\label{Eq_SystemF}
\Bold{q}&=[q_j]_{1 \times l}, ~~ 0 \leq q_j \leq 1 \\
\Bold{\RV{Q}}&=[\RV{Q}_j]_{1 \times l}
\end{align}

Since $q_j$ represents failure, the success of subsystem $j$ is $1 - q_j$. Assuming independence between subsystems, the success of function $k$ can be defined as:
\begin{align}
&s_k=\prod_{\substack{j=1,\\f_{jk}=1}}^{l} (1-q_j)=\prod_{j=1}^{l} (1-q_j)^{f_{jk}} \label{Eq_SuccFunc}
\end{align}

As defined in \Eq{Eq_MapF}, $f_{jk}$ indicates whether subsystem $j$ supports function $k$. The $f_{jk}$ element in \Eq{Eq_SuccFunc} is used to filter the relevant subsystems. If all the relevant subsystems are fully functional ($q_j=0$), the function succeeds completely ($s_k=1$). If one associated subsystem fails completely ($q_j=1$), the function fails completely ($s_k=0$). In cases of partial success of the subsystems ($0 < q_j < 1$), the function achieves partial success ($0 < s_k < 1$).

\section{Mitigation targets}\label{Sec_MitigTargets}

In the previous three sections, we established the foundational principles relating to the success and failure of mitigation functions and their role in reducing risk. This section brings those principles together into a modelling approach that enables estimation of the target risk reduction to be achieved by a set of mitigation subsystems -- analogous to the conventional SIL determination process.

In simple terms, our approach is as follows: risk depends on the expected consequence severity, which in turn depends on the expected success/failure of the mitigation functions, which themselves depend on the expected success/failure of the associated subsystems. If we can define the consequence $c$ in terms of subsystem performance, we can derive reliability targets for the subsystems such that, if met, the overall expected risk remains below the tolerable risk threshold.

We define the consequence $c$ as a function of the success levels $\Bold{s}$ and contribution levels $\Bold{u}$:
\begin{align}
c = &c_{max} - (c_{max} - c_{min}) \sum_{k=1}^{m} (u_k \cdot s_k) \label{Eq_Cons1}
\end{align}

This expression implies that consequence severity ranges from $c_{min}$ to $c_{max}$, reaching its minimum when all functions succeed completely ($s_k = 1$ for all $k$s), and its maximum when all functions fail completely ($s_k = 0$ for all $k$s).

Substituting the expression for $s_k$ from \Eq{Eq_SuccFunc} into \Eq{Eq_Cons1}, we can express $c$ directly as a function of subsystem failures $\Bold{q}$:\footnote{Note that \Eq{Eq_Cons2} is just one possible formulation of the consequence as a function of subsystem failures. In specific applications, the analyst may adopt a different expression to reflect the system’s context. Nonetheless, the requirement is to be able to mathematically express $c$ as a function of $q_j$s.}
\begin{align}
c = &c_{max} - (c_{max} - c_{min}) \sum_{k=1}^{m} ( u_k \prod_{j=1}^{l} (1 - q_j)^{f_{jk}} ) \label{Eq_Cons2}
\end{align}

Given that $c$ is now expressed as a function of the random variables $\Bold{q}$, we can derive the expected consequence severity $E[\RV{C}]$ in terms of the expected failure degrees $E[q_j]$. Assuming independence among subsystems:
\begin{align}
E[\RV{C}] = c_{max} - (c_{max} - c_{min}) \sum_{k=1}^{m} ( u_k \prod_{j=1}^{l} (1 - E[q_j])^{f_{jk}} ) \label{Eq_ExpConsLast}
\end{align}

To satisfy the tolerable consequence constraint expressed in \Eq{Eq_RiskTolConseq}, the following inequality must hold:
\begin{align}
c_{max} - (c_{max} - c_{min}) \sum_{k=1}^{m} ( u_k \prod_{j=1}^{l} (1 - E[q_j])^{f_{jk}} ) \leq \overline{E[\RV{C}]} \label{Eq_ExpConsLast1}
\end{align}

Thus, the SIL allocation problem is redefined as follows: determine the set of expected subsystem failure values $E[q_j]$ such that the resulting expected consequence severity $E[\RV{C}]$ ensures that the estimated risk $r$ calculated by \Eq{Eq_RiskDiscB11} does not exceed the tolerable risk $\overline{r}$. These $E[q_j]$ values then become the design targets for subsystem reliability.

In simple cases with few subsystems, $E[q_j]$ targets can be identified through manual calculation. In more complex scenarios, numerical optimisation methods may be employed. If used, these methods can also incorporate implementation costs, allowing trade-off analyses between risk reduction and cost-efficiency.

Unlike the single probability value of PFD, which is often averaged over the mission time of the SF, the expected value $E[q_j]$ represents the behaviour of the PDF function $f_{\RV{Q}_j}(q_j)$ over the full range of subsystem's failure $q_j \in [0, 1]$. This is particularly relevant in mitigation functions, where the risk is affected by the degree of consequence severity.

As discussed earlier, expected values $E[q_j]$ can also be defined and applied to binary systems. Therefore, the approach presented here may also be extended to the study of preventive systems, where binary failure models are prevalent. Our method is not intended as a replacement for conventional binary approaches, such as LOPA, but rather as a generalisation of these methods -- to address the analysis of both binary and non-binary systems in one framework.

~

In summary, allocating target expected failure values $E[q_j]$ to subsystems follows these steps:
\begin{enumerate}[left=0.5cm]
    \item Identify the frequency of the initiating event $w_{HE}$.
    \item Identify the functions and subsystems.
    \item Construct the mapping matrix $\Bold{F}$.
    \item Define the consequence segments and assign values $c_h$.
    \item Determine $c_{max}$ and $c_{min}$ based on $c_h$ values.
    \item Establish the tolerable level of risk $\overline{r}$.
    \item Define the contribution values $u_k$ and build vector $\Bold{u}$.
    \item Use \Eq{Eq_RiskDiscB111} to calculate $\overline{E[\RV{C}]}$ based on $\overline{r}$ and $w_{HE}$.
    \item Use \Eq{Eq_ExpConsLast1} to compute suitable $E[q_j]$ values.
    \item Derive design requirements from the resulting values $E[q_j]$.
\end{enumerate}

\section{Final thoughts}\label{Sec_FinalPoints}

\subsubsection{All random variables:}
Consequence severity is a random variable that depends on $c_{max}$, $c_{min}$, $\Bold{u}$, and $\Bold{q}$, as shown in \Eq{Eq_Cons2}. In general, all of these variables are themselves random because every accident unfolds differently.

As an example, $c_{\max}$ depends on the initial extent of the hazardous event that needs to be mitigated, and it may vary widely. For instance, in a road tunnel fire scenario, the size of a fire breaking out from a passenger car differs from one breaking out from a fuel tanker; hence the assignment of different $c_{\max}$ values.

However, in analyses such as SIL allocation, where the focus is on system-related targets, these variables can be approximated by constant values, as we did in \Eq{Eq_Cons1} - \Eq{Eq_ExpConsLast}. Non-system variables, such as $c_{\max}$, may be studied more closely in other analyses, where both system and non-system factors need to be included for accident modelling, for instance.

\subsubsection{Pre-accident mitigation:}
Mitigation functions typically come into action after the occurrence of a hazardous event -- like a fire response system that is activated after a fire breaks out. However, there are systems that can reduce risk by reducing the \emph{potential} consequences of an accident -- before it occurs.

Consider, for instance, a system that detects the leakage of a flammable gas and activates a ventilation system to reduce the concentration level. Any reduction in concentration can reduce the severity of consequences if ignition occurs. The ventilation function works prior to the fire event to reduce its consequences -- if the fire happens at all.

From another perspective, we can consider the ventilation function a post-event function if we define the loss of containment (i.e., the leakage) as the primary hazardous event.

\subsubsection{Hazardous events:}
The Bowtie model shown earlier in \Fig{Fig_Bowtie} is, obviously, a simplified demonstration. In realistic scenarios, an accident may unfold in multiple stages, where each event can be viewed as a hazardous event caused by another hazardous event. For example, in a house fire caused by a gas leak, a faulty pipe may initiate the sequence, causing a gas leakage as the first hazardous event (HE1), which may lead to fire and explosion as a second hazardous event (HE2), which in turn may result in the collapse of the building (HE3). Each of these HEs has its own potential consequences, while some (HE1 and HE2) also initiate subsequent HEs.

In general, the alignment of initiating events, hazardous events, and Safety Function(s) can vary based on the scope and objectives of the analysis. What matters most is that the overall model is sound, irrespective of the names and labels, and that the model serves the objectives of the analysis.

This is particularly important because the position of ``initiating events'' and ``hazardous events'' can change the characterisation of the Safety Function. A function that mitigates the consequences of HE1 may be considered a preventive function with respect to HE2. However, in general, and as defined earlier in \Sec{Sec_RiskMitig}, if a function reduces risk by reducing consequences -- within the context of the analysis -- it is considered a mitigation function.

\subsubsection{External influences:}
In safety analyses such as SIL allocation, the focus is on the system. However, accidents may also depend on other random factors that are not related to the system. Recall the gas ventilation example. The ventilation system can only reduce the level of gas concentration. However, for a fire accident to occur, an ignition source is also required -- which is an external element to the system. That is, the failure of a Safety Function, preventive or mitigation, may not directly lead to an accident, and the role of external factors should be considered too.

In SIL allocation studies, where the focus is on the role of the system, external enablers and triggers such as ignition are typically factored in using constant probability and frequency values, which may require adjustments to the generic model we introduced in \Sec{Sec_MitigTargets}. 

\subsubsection{Double-role functions:}
Classifying a function as mitigation is not always straightforward. There are situations where a system reduces risk by both lowering the probability of an accident and lessening the severity of its consequences. Consider, for example, a hydrocarbon detection and ventilation system. Hydrocarbon accumulation can result in fatal accidents if a minimum level of gas concentration (LEL) finds a source of ignition. An effective detection and ventilation system reduces the probability of accident by keeping the concentration below the LEL, but it can also reduce the severity of the explosion if it only reduces the concentration to a level that is still above the LEL.

In general, where the preventive portion is negligible, we consider the function to be solely mitigation. Similarly, if the mitigation portion is negligible, we only factor it in as a preventive function. Where the two portions are comparable, one option may be to break the function into two sub-functions -- one for the prevention part and one for the mitigation part. Where that is not practical, the risk scenario can be divided into two or more separate scenarios to cover all possible outcomes. Alternatively, where the same function is implemented and the distinction between prevention and mitigation is determined by other (external) factors, setting $c_{min}=\V{0.0}$ can help cater for the prevention portion. An example of this approach is given in \Sec{Sec_FlammableGas}.

\subsubsection{The fourth element:}

In its simplest form, risk is a combination of the likelihood of occurrence ($w_{HE}$) and the severity of consequence ($c_{HE}$). We also know that both $w_{HE}$ and $c_{HE}$ are random variables, influenced by the likelihood of initiating events, the success/failure of preventive functions, and the performance of mitigation functions. 

There is, however, a fourth element that is often overlooked in mitigation safety analyses for the sake of simplification: the initial severity of the HE, to which mitigation functions are intended to respond. This element often depends on other (external) factors (e.g., the type of vehicle involved in a road tunnel fire scenario). By simply stating that the frequency of the hazardous event is $w_{HE}$, we implicitly assume that the initial scale of the incident is identical in every occurrence. This assumption, however, is rarely valid in real-world scenarios. 

A more accurate approach would be to use the expected values to determine the ``centre of gravity'' of initial severity. This expected severity can then be incorporated into the analysis, either by adjusting $c_{min}$ and $c_{max}$, or by introducing a separate external parameter in the model.

\section{Case studies}\label{Sec_CaseStudy}

In this section, we demonstrate the implementation of our method through two case studies.  

\subsection{Flammable gas scenario}\label{Sec_FlammableGas}

\begin{figure}[!h]
\begin{center}
\includegraphics[scale=0.5]{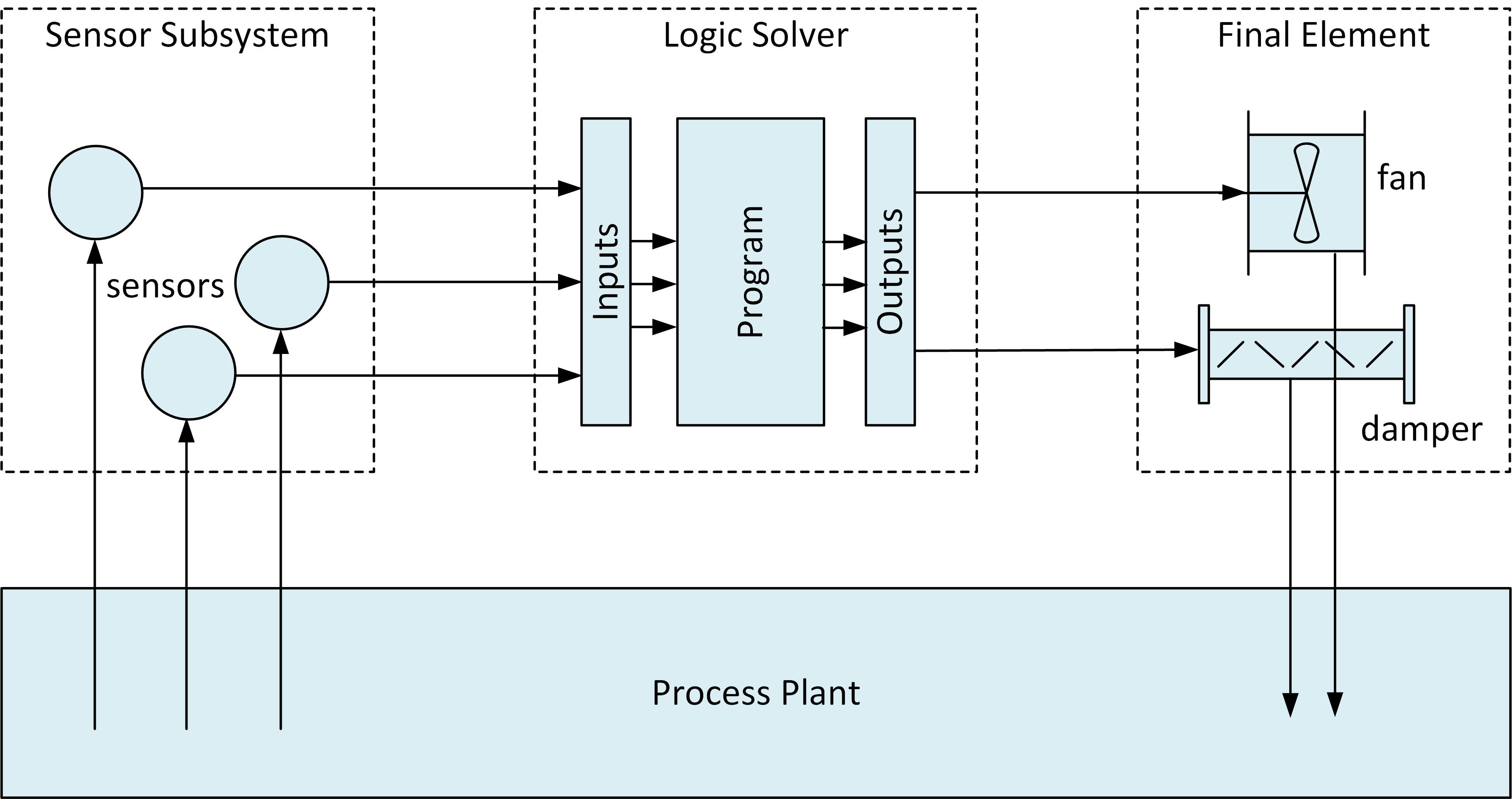}
\end{center}
\caption{Hydrocarbon Ventilation System}
\label{Fig_VentSystem}
\end{figure}

Consider a hydrocarbon ventilation system comprising a sensor subsystem that monitors the concentration of flammable gas; a logic solver subsystem  that interprets the sensor readings and determines whether the concentration has reached a dangerous level; and a final element subsystem, consisting of an extraction fan that vents the gas into the atmosphere and a multi-module damper that isolates the fan from the process. The system configuration is illustrated in \Fig{Fig_VentSystem}.

Hydrocarbon is used in oil and gas processes. Its accumulation can result in fire and explosion accidents if a minimum level of gas concentration (LEL) encounters a source of ignition. Piper Alpha \cite{Ref_381} is one example of such accidents.

The following aspects of this case study are notable with respect to the points made in \Sec{Sec_FinalPoints}:

\begin{itemize}
\item The ventilation function serves two purposes: it can \emph{prevent} explosions by reducing the gas concentration below the LEL, or \emph{mitigate} the consequences if that threshold is exceeded and a fire breaks out. Since both outcomes are achieved by the same function, we set $c_{min} = \V{0.0}$ to represent the case where the explosion is prevented and no harm occurs.

\item This is a case where the mitigation function is performed \emph{prior} to the accident, as the explosion also requires a source of ignition.
\item The occurrence of an accident depends on an \emph{external} factor -- namely the ignition source -- which is independent of the performance of the mitigation system. In our case study, this external factor is treated as an additional probability term in the risk equation.
\item This case study includes both binary and non-binary subsystems, demonstrating how our PDF-based method can address both types.
\end{itemize}

Following the steps listed in \Sec{Sec_MitigTargets}, we begin by identifying the baseline data. Suppose the frequency of a hydrocarbon leakage is 1 in 10 years; that is, $w_{HE} = \V{0.1}$ per year. Let us also assume that the probability of ignition in the area is ten percent: $p_{ign} = \V{0.1}$. Thus, the risk equation presented in \Eq{Eq_RiskDiscB11} is modified to the following form:
\begin{align}
&r = w_{HE} \cdot p_{ign} \cdot E[\RV{C}] = \V{0.01} E[\RV{C}] \label{Eq_RiskDiscB112}
\end{align}
We also modify \Eq{Eq_RiskDiscB111} to incorporate $p_{ign}$:
\begin{align}
&\overline{E[\RV{C}]} = \overline{r} / (w_{HE} \cdot p_{ign}) \label{Eq_RiskDiscB113}
\end{align}

Here, there is only one function and three subsystems involved. Hence, $m = 1$, $l = 3$, and the contribution vector is $\Bold{u} = [1]$. Since the function depends on all three subsystems, the mapping matrix $\Bold{F}$ is a single-column vector with $f_{jk} = 1$ for $1 \leq j \leq 3$ and $k = 1$.

In this study, the severity of consequences is expressed in monetary terms, assuming that human casualties are converted into financial values and included. Suppose $c_{max} = \$\V{250M}$ and, as noted earlier, $c_{min} = \$\V{0}$ to account for the preventive impact. 

Let the risk tolerance threshold be set at $\overline{r} = \$\V{0.1M}$, which, from \Eq{Eq_RiskDiscB113}, gives:
\begin{align}
&\overline{E[\RV{C}]} = \$\V{10M} 
\end{align}
The goal is to find the $E[q_j]$ values such that $r \leq \overline{r}$, or $E[\RV{C}] \leq \overline{E[\RV{C}]}$. Using \Eq{Eq_ExpConsLast1}, the risk constraint becomes:
\begin{align}
&c_{max} ( 1 - \prod_{j=1}^{3} (1 - E[q_j])) \leq \overline{E[\RV{C}]} \implies \nonumber \\
&\V{250}  ( 1 - \prod_{j=1}^{3} (1 - E[q_j])) \leq \V{10} \implies \nonumber \\
&(1 - E[q_s])(1 - E[q_{ls}])(1 - E[q_{fe}]) \geq \V{0.96} \label{Eq_SubsRiskInequality}
\end{align}
where $q_s$, $q_{ls}$, and $q_{fe}$ represent the degree of failure of the sensor, logic solver, and final element subsystems, respectively.

A conventional SIL allocation would proceed by assigning a combined target PFD across all three subsystems. Below, we illustrate how using expected degrees of failure -- rather than binary PFD targets -- can lead to more meaningful and practical design decisions.

Consider the sensor subsystem first. The performance of a sensor depends on its accuracy and its response time. These qualities can be influenced by many factors, which we do not intend to cover in this paper. However, one can observe that accuracy and response time are inherently non-binary characteristics. A late detection of gas is not a complete failure; it is a partial one, with a variable impact depending on the delay. 

Let $T_{max}$ be the time after which a gas leak may result in maximum consequences if not detected. If $T$ is the sensor’s actual response time, then one simple definition for its degree of failure could be $q_s = T / T_{max}$, which means its expected success is:
\begin{align}
&1 - E[q_s] = 1 - E[T]/T_{max} \label{Eq_FailSens}
\end{align}

This offers system designers a clearer target: a required expected response time $E[T]$, rather than a probability value that is, first, allocated to the entire SF rather than just the sensor subsystem, and, second, does not accurately represent the proportional performance of the component.

The logic solver, on the other hand, is a binary component. As shown earlier in \Eq{Eq_ExpProb}, its expected success is equal to its probability of success:
\begin{align}
&1 - E[q_{ls}] = 1 - p_{ls} \label{Eq_FailLS}
\end{align}
where $p_{ls}$ is the logic solver’s PFD.

The final element comprises two parts: a fan (binary) and a multi-module damper (non-binary). Similar to the binary logic solver, the success of the fan can be expressed by:
\begin{align}
&1 - E[q_{fan}] = 1 - p_{fan} \label{Eq_FailFan}
\end{align}
where $p_{fan}$ is the fan’s PFD.

For the damper, however, we take an approach similar to the non-binary sensor. Let $O_{max}$ be the total number of damper modules, and $E[O]$ the expected number that succeed in opening. Then one definition for the degree of success of the damper can be:
\begin{align}
&1 - E[q_{dmp}] = E[O]/O_{max} \label{Eq_FailDamper}
\end{align}

Combining \Eq{Eq_FailFan} and \Eq{Eq_FailDamper}, we define the degree of success for the final element:
\begin{align}
&1 - E[q_{fe}] = (1 - p_{fan})(E[O]/O_{max}) \label{Eq_FailFE}
\end{align}

Substituting \Eq{Eq_FailSens}, \Eq{Eq_FailLS}, and \Eq{Eq_FailFE} into \Eq{Eq_SubsRiskInequality}, we get:
\begin{align}
&(1 - E[T]/T_{max})(1 - p_{ls})(1 - p_{fan})(E[O]/O_{max}) \geq \V{0.96} \label{Eq_SubsRiskInequality1}
\end{align}

The task is now to determine suitable values for $E[T]$, $p_{ls}$, $p_{fan}$, and $E[O]$ such that \Eq{Eq_SubsRiskInequality1} holds.

As far as the risk constraint is concerned, any combination of values that meets this inequality is acceptable. This allows cost and feasibility to drive the design. Formally, this can be framed as a cost-minimisation problem with \Eq{Eq_SubsRiskInequality1} as a constraint. While solving such optimisation problems could be a future research direction, in this case study we used manual trial and error guided by expert judgment. The final chosen values are:
\[
\begin{array}{cc}
E[T] = \V{3.4}\%~\text{of}~T_{max}~~~~~~ & ~~~~~~p_{ls} = \V{1.0E-04} \\
p_{fan} = \V{2.0E-03}~~~~~~ & ~~~~~~E[O] = \V{99.6}\%~\text{of}~O_{max} 
\end{array}
\] 

Substituting these values into \Eq{Eq_SubsRiskInequality1}, the left-hand side evaluates to $\V{0.96}$, meeting the required thresholds of $\overline{r}$ and $\overline{E[C]}$.

Note that for the sensor subsystem, only the expected response time is set -- not the number of sensors. As we discuss later in \Sec{Sec_Discussion}, random failure is only one element of safety integrity. The system architecture must also meet the requirements of Hardware Fault Tolerance, which dictate the number and voting configuration of the sensors.

Similarly, for the dampers, the design target is to achieve an expected opening of $\V{99.6}\%$. This, however, does not specify the total number of modules. That can be determined based on other factors, such as cost and construction. In terms of the probability of failure, the damper follows a binomial distribution \cite{Ref_204}, for which the expected value of opening is $E[O] = O_{max}(1 - p_{mod})$, with $p_{mod}$ being the probability of one module failing to open. Therefore, irrespective of the total number of modules, the target PFD for one module will always be $p_{mod} = \V{4.0E-03}$.

\subsection{Road tunnel fire scenario}\label{Sec_Case1}

For the second case study, we use the fire road tunnel fire scenario presented in \cite{Ref_382}.\footnote{Readers are encouraged to refer to \cite{Ref_382} for a detailed description of the case study.} This scenario concerns the fire incidents in a 6\,km twin road tunnel with a frequency of occurrence of $w_{HE}=\V{0.7}$ per year. The functions triggered during fire events are:
\begin{itemize}
\item AFS: Automatic Fire Suppression
\item MFS: Manual Fire Suppression
\item ASE: Automatic Smoke Extraction
\item MSE: Manual Smoke Extraction
\item EE: Emergency Evacuation
\end{itemize}

Among these, ASE is the function for which SIL allocation is considered. These functions are supported by the following subsystems:
\begin{itemize}
\item LHD: Line Heat Detector -- detects fire by sensing heat
\item FDP: Fire Detection Panel -- interfaces between the LHD and other systems
\item IAD: Intelligent Accident Detector -- includes cameras and image-processing algorithms capable of alarming on fire incidents
\item PCS: Plant Control System
\item TOp: Tunnel (human) Operator
\item OMS: Operation Management System -- interfaces between TOp and PCS
\item FSS: Fire Suppression System -- includes deluge valves
\item TVS: Tunnel Ventilation System -- extracts smoke from the tunnel
\item EMS: Egress Management System -- includes emergency lighting and signage 
\item TUs: Tunnel Users (i.e., drivers and passengers)
\end{itemize}

With $m=\V{5}$ functions and $l=\V{10}$ subsystems, the mapping matrix $\Bold{F}$ was defined as \cite{Ref_382}:
\[
 \newcommand\col[1]{\parbox{1cm}{\centering\scriptsize#1}}
 \Bold{F} = 
 \begin{blockarray}{cccccc}
 AFS ~&~ MFS ~&~ ASE ~&~ MSE ~&~ ~EE ~&~ \\
 \begin{block}{[ccccc]l}
 1 ~&~ 0 ~&~ 1 ~&~ 0 ~&~ 0 ~&~ ~LHD \\
 1 ~&~ 1 ~&~ 1 ~&~ 0 ~&~ 0 ~&~ ~FDP \\
 0 ~&~ 1 ~&~ 0 ~&~ 0 ~&~ 0 ~&~ ~IAD \\
 0 ~&~ 1 ~&~ 1 ~&~ 1 ~&~ 1 ~&~ ~PCS \\
 0 ~&~ 1 ~&~ 0 ~&~ 1 ~&~ 1 ~&~ ~TOp \\
 0 ~&~ 1 ~&~ 0 ~&~ 1 ~&~ 1 ~&~ ~OMS \\
 1 ~&~ 1 ~&~ 0 ~&~ 0 ~&~ 0 ~&~ ~FSS \\
 0 ~&~ 0 ~&~ 1 ~&~ 1 ~&~ 0 ~&~ ~TVS \\
 0 ~&~ 0 ~&~ 0 ~&~ 0 ~&~ 1 ~&~ ~EMS \\
 0 ~&~ 0 ~&~ 0 ~&~ 0 ~&~ 1 ~&~ ~TUs \\
 \end{block}
 \end{blockarray}
\]

The risk criteria used in this study are presented in \Table{Table_Cons}, where the first column denotes the severity segments, the last column specifies their tolerable frequencies, and the two middle columns show the severity levels.

\begin{table}[htbp]
\begin{center}
\begin{tabular}{|l|l|l|l|}
\hline \multicolumn{1}{|c|}{\TE{Conseq. Seg.}}&\multicolumn{1}{c|}{\TE{Safety}}&\multicolumn{1}{c|}{\TE{Financial}}&\multicolumn{1}{c|}{\TE{Tolerance Freq.}}\\
\specialrule{0.2em}{0.0em}{0.0em}
\TE{Catastrophic} & \TE{Fatalities} & \TE{$\geq \V{\$40M}$} & \TE{0.001 p.y.} \\
\TE{Major} & \TE{Permanent disabilities} & \TE{$\V{\$20} - \V{\$40M}$} & \TE{0.01 p.y.} \\
\TE{Moderate} & \TE{Hospitalisation} & \TE{$\V{\$5} - \V{\$20M}$} & \TE{0.1 p.y.} \\
\TE{Minor} & \TE{Medical treatments} & \TE{$\V{\$0.5M} - \V{\$5M}$} & \TE{1 p.y.} \\
\TE{Insignificant} & \TE{First aid} & \TE{$\leq \V{\$0.5M}$} & \TE{10 p.y.} \\
\hline
\end{tabular}
\end{center}
\caption{Risk Tolerance Criteria}
\label{Table_Cons}
\end{table}

As mentioned earlier, the tolerable risk $\overline{r}$ is driven by policy and compliance requirements. However, for the purpose of this case study -- and given the information in \Table{Table_Cons} -- we employ \Eq{Eq_RsikPDF_Tol} to determine $\overline{r}$ based on the values of $c_h$ and $\overline{w}_h$.

Safety consequences can be translated into equivalent values using the Fatalities and Weighted Injuries (FWI) metric. The ratios between different categories of casualties are typically defined by national authorities \cite{Ref_378,Ref_379,Ref_380}. To closely align with the approach used in \cite{Ref_378}, which is used in Australia, we adopt the following equivalency ratios: 1 fatality = 3 permanent disabilities = 10 hospitalisations = 200 medical treatments = 1000 first aid cases. Thus, the severity levels are quantified as:

\begin{itemize}
  \item Catastrophic: $c_1 = \V{1.0}$
  \item Major: $c_2 = \V{0.333}$
  \item Moderate: $c_3 = \V{0.1}$
  \item Minor: $c_4 = \V{0.005}$
  \item Insignificant: $c_5 = \V{0.001}$
\end{itemize}

By substituting these values, along with the $\overline{w}_h$ values from \Table{Table_Cons}, into \Eq{Eq_RsikPDF_Tol}, the tolerable risk $\overline{r}$ is calculated as:
\begin{align}
\overline{r} &= \V{0.0293} \nonumber
\end{align}

In this case study, consequences are expressed in terms of fatalities. While it is possible to aggregate safety and financial risks into a unified scale (e.g., monetary or utility values) \cite{Ref_377,Ref_345,Ref_346}, we avoid aggregation here for simplicity. Consequently, the third column in \Table{Table_Cons} is not utilized. 

The highest severity level in \Table{Table_Cons} is ``Catastrophic'', which begins at a single fatality but may exceed this. To account for the possibility of multiple casualties, let us assume $c_{max} = \V{2.0}$.\footnote{A fire incident may result in a higher number of casualties. However, in the context of SIL allocation, only the system-related portion of risk may be considered -- rather than the total number of casualties -- since the baseline casualty rate, such as that on open roads, is not attributable to the tunnel system.} For simplicity, we set $c_{min} = \V{0.0}$. 

We use \Eq{Eq_RiskDiscB111} to set the tolerable expected consequence severity $\overline{E[\RV{C}]}$ based on $\overline{r}$ and $w_{HE}$:
\begin{align}
\overline{E[\RV{C}]} &= \V{0.0419} \nonumber
\end{align}

Note that, as mentioned at the end of \Sec{Sec_ExpRisk}, the tolerance threshold $\overline{r}$ represents the collective value of risk across all consequence segments and should not be confused with the tolerance thresholds for individual segments. Similarly, $\overline{E[\RV{C}]}$ does not imply that we tolerate $\V{0.0419}$ fatalities per accident; because $\overline{E[\RV{C}]}$ is calculated based on the collective value of $\overline{r}$.

To estimate the actual expected consequence severity $E[\RV{C}]$, we define the contribution vector $\Bold{u}$. Suppose the analysis team agrees -- based on subject-matter expertise -- that the contributions of the individual functions to mitigating the fire event are:
\begin{align}
\Bold{u} = [\V{0.15} ~~ \V{0.15} ~~ \V{0.3} ~~ \V{0.2} ~~ \V{0.2}] \label{Eq_Contrib}
\end{align}
That is, the functions AFS, MFS, ASE, MSE, and EE contribute $\V{15\%}$, $\V{15\%}$, $\V{30\%}$, $\V{20\%}$, and $\V{20\%}$, respectively, to fire and smoke mitigation.

The objective now is to define $E[q]$ such that the calculated value of $E[\RV{C}]$ remains below $\overline{E[\RV{C}]}$ as per \Eq{Eq_ExpConsLast1}. A trial-and-error approach was employed, incorporating expert judgement regarding the reliability ratios between subsystems, which yielded:

\begin{align}
E[\Bold{q}] = \nonumber [&\V{1.2E-04} ~~ \V{9.0E-05} ~~ \V{1.0E-02} ~~ \nonumber \\
&\V{9.0E-05} ~~ \V{1.4E-02} ~~ \V{2.0E-03} ~~ \V{2.0E-03} ~~ \V{2.0E-04} ~~ \V{1.4E-02} ~~ \V{3.6E-02}] \label{Eq_Eqj}
\end{align}

While the values given in \Eq{Eq_Eqj} set targets for all subsystems, the Safety Function ASE only concerns the reliability targets of its own subsystems: LHD, FDP, PCS, and TVS:
\begin{align}
E[q_1] = \V{1.2E-04}, \quad E[q_2] = \V{9.0E-05}, \quad E[q_4] = \V{9.0E-05}, \quad E[q_8] = \V{2.0E-04} \nonumber
\end{align}

Together with the other values from \Eq{Eq_Eqj}, these meet the boundary risk limit of $\overline{r} = \V{0.0293}$ and can therefore be adopted as the target expected degrees of failure for their respective subsystems. These $E[q_j]$ values can now be translated into practical design requirements, in a manner similar to that outlined in \Sec{Sec_FlammableGas}.

In \cite{Ref_382}, the trial-and-error process resulted in a collective target PFD. The $E[q_j]$ values calculated here are not probability values, even though they lie between 0 and 1. They represent the target expected degree of failure -- the centre of gravity across the full range of performance. In contrast, the PFD values calculated in \cite{Ref_382} reflect a binary distinction between success and failure. That is, the SIL allocation process in \cite{Ref_382} relied on the key assumption that the non-binary performance of the mitigation function could be reduced to a binary state, primarily for the sake of simplification. In the PDF-approach, such assumptions and reductions are not necessary.

Furthermore, as in other typical SIL allocation methods, the case study in \cite{Ref_382} yielded a collective PFD for the entire SF. In contrast, the approach presented in this paper assigns separate targets to individual subsystems. While IEC 61508 does not explicitly require the apportionment of the target PFD to subsystems, the targets derived from our method relate directly to subsystem-level performance. Since these targets may carry different meanings across subsystems, assigning individual targets allows each one to be translated into practical and meaningful design requirements specific to the respective subsystem.

\section{Discussion}\label{Sec_Discussion}

Preventive and mitigation Safety Functions differ from various perspectives. We previously highlighted one main difference: the characterisation of failure. Here are some more differences worth noting (also see \cite{Ref_368} and \cite{Ref_369}):

\begin{itemize}
\item \textbf{The primary element of risk control:} Preventive SFs reduce risk by lowering the probability of an event, while mitigation SFs reduce risk by lessening the severity of its consequences. Although, the impact on severity, too, is probabilistic. 

\item \textbf{Event frequencies:} In prevention, we only compare the frequency of the hazardous event with a tolerance threshold frequency. In mitigation, we need to examine the tolerance of frequencies across a range of consequence severity segments, while summarising all occurrences of the hazardous event into one frequency element, $w_{HE}$.

\item \textbf{Reliability judgement:} The reliability of a preventive SF is measured by the probability of being in a failed or successful state. The reliability of a mitigation SF can be measured by its probabilistic performance in reducing the consequence severity.

\item \textbf{Level of consequence:} In mitigation SFs, the severity of consequence is the dominating element that determines the risk. In preventive SFs, consequence severity is not included in calculating the risk gap.

\item \textbf{Sequence of events:} In prevention analyses, the hazard scenario begins with multiple events and concludes with one (i.e., the HE). In mitigation scenarios, it starts with one event and results in multiple possible outcomes.
\end{itemize}

The inherent differences between the two types, particularly the non-binary success/failure state in mitigation SFs, demand different approaches to setting target reliability measures. This is why we proposed a conceptually different approach based on PDF and expected values. 

In the remainder of this section, we briefly discuss some other aspects of PDF-based approach. 
 
\subsubsection{Deriving SIL from PDF:}

Unlike the probability of failure, which is a measure rooted in the scientific principles of probability theory, SIL is a measure built on consensus and agreement. It is the industry community that has devised the discrete scale of SIL and linked it to PFD/PFH. Non-probability measures, such as PDF and expected value proposed here, should not be expected to be directly convertible to SIL. 

One indirect way to address this gap is to assess the impact of the SF on the overall risk, as also suggested by the standard in Clause A.9 of \cite{Ref_187}; see the excerpt provided earlier in \Sec{Sec_Background}. 

In the hydrocarbon case study, for example, the initial risk (i.e., in the absence of the SF) is $r_{max} = w_{HE} \cdot p_{ign} \cdot c_{max} = \$\V{2.5M}$. Assuming that the SF meets its targets, this risk will be reduced to the tolerable level of $\overline{r} = \$\V{0.1M}$. Therefore, the SF provides a Risk Reduction Factor of $RRF = \V{25}$, which corresponds to SIL1 according to the tables provided in the standard \cite{Ref_184}.

However, this approach may not always be straightforward. For example, in the road tunnel case discussed in \Sec{Sec_Case1}, estimating the risk in the absence of the ASE function is challenging due to its shared subsystems with other functions. In such cases, assigning SIL at the subsystem level -- rather than at the function level -- may be more appropriate, as it can help disentangle interdependent functions. However, this concept requires further research and lies beyond the scope of this paper.    

\subsubsection{Hardware Fault Tolerance (HFT):}

Parallel to PFD, which measures the level of hardware random failure, the functional safety standard IEC 61508 uses the discrete measure HFT to gauge the architectural integrity of the safety system. In a subsystem with a number of identical redundant channels, HFT is the number of failed channels that the subsystem can tolerate without failing the SF. In a 3oo4 subsystem, for instance, HFT is 1 because the failure of more than 1 channel results in the failure of the whole subsystem. 

This current definition of HFT is not applicable to mitigation subsystems where success and failure are assessed on a proportional scale. In such cases, any level of failure concurrently represents a degree of success. Therefore, it becomes impractical to assign a discrete numerical value to HFT that would represent a distinct boundary between success and failure. 

In a generalised sense, HFT can be formulated as the \emph{degree of degradation} that can be tolerated from one fault. Hence, in its general form, HFT does not need to be a discrete number. An alternative that can work for both binary and non-binary systems is a percentage scale. In that case, a target HFT of a mitigation subsystem could indicate the extent of success compromised by the event of one element's complete failure. This definition closely aligns with the current definition of HFT (for preventive SFs) and could be used to assure architectural integrity in mitigation subsystems. As this may require further effort to establish a solid concept, we leave this topic for another research opportunity.

\subsubsection{Proof Test Interval (PTI):}

PFD is directly linked to the test interval. For a single component: $PFD_{avg} = \lambda_{DU} \cdot \tau / 2$, in which $PFD_{avg}$ is the average PFD, $\lambda_{DU}$ is the Dangerous Undetected failure rate, and $\tau$ is the PTI.

When taking the PDF-based approach, a different method should be used to determine PTI. One possible research direction is to define a degradation rate function $d(t)$, where the cumulative expected failure grows over time, and then assess how test intervals truncate this growth. Integrating $d(t)$ over each PTI cycle could yield an equivalent measure analogous to average PFD, but interpreted in terms of accumulated expected failure rather than binary failure probability.

In summary, while the necessity for periodic testing remains, the method and interpretation of PTI must be revisited under the PDF paradigm. Further research is required to formalise how PTI influences the evolving distribution of functional performance, and to define test coverage requirements.

\subsubsection{Optimisation and the impact of cost:}

A body of work has focused on optimisation methods used in the context of risk assessment and risk management \cite{Ref_345,Ref_370,Ref_371,Ref_372,Ref_373}. In general, the main objective of these studies is to find a set of parameters that minimise a target criterion (e.g., a combination of risk and cost). In the context of our work, this relates to the problem of apportioning the target reliability values $E[q_j]$ among multiple subsystems such that a risk limitation is met and the cost is minimised. However, the focus of our paper was on the underlying concepts and formulation. The impact of cost and the optimisation of target allocation can be addressed in future work.

\section{Conclusion}\label{Sec_Conclusion}

In this paper, we argued that PFD may not be the optimal metric for expressing the reliability of mitigation functions. Instead, we proposed a new measure based on the Probability Density Function (PDF) and the expected degree of failure. We formulated a PDF-based methodology, supported by two practical case studies that demonstrated its realistic application.

The core challenges addressed in this study stem from the inherent differences between mitigation and preventive functions. These include distinctions in the definitions of success and failure, variations in risk reduction mechanisms, and the complexities introduced by common-cause failures and subsystem dependencies.

The PDF-based approach introduced here offers a new perspective -- one that accounts for the probabilistic nature of failure across the full spectrum of functional performance. We believe this method provides a more realistic and informative representation of failure for mitigation functions, compared to the conventional use of single PFD values. However, we acknowledge that the increased accuracy and insight come at the cost of slightly more complex calculations. Additionally, the development of a fully standardised PDF-based methodology will require further deliberation and consensus within industry standardisation bodies before it can be adopted in practice.

In closing, we suggest the following directions for future research: (i) developing a method to relate the expected degree of failure to SIL, (ii) defining Hardware Fault Tolerance for non-binary subsystems, (iii) establishing appropriate Proof Test Intervals when applying the PDF-based approach, and (iv) applying numerical optimisation techniques to allocate target probability parameters, considering the cost of implementation.

\bibliographystyle{ieeetr} 
\balance
\bibliography{References}

\end{document}

%% file: Commands.TEX
\usepackage{graphicx}
\usepackage{longtable}
\usepackage{amsfonts}
\usepackage{amssymb}
\usepackage{bm}
\usepackage[utf8]{inputenc}
\usepackage[english]{babel}
\usepackage{xstring}
\usepackage{setspace}
\usepackage{array}
\usepackage{subfig}
\usepackage{enumitem}
\usepackage{multirow}
\usepackage{algorithmic}
\usepackage{dutchcal} 			
\usepackage{ctable} 				
\usepackage{comment}
\usepackage[ruled,vlined,linesnumbered]{algorithm2e}
\usepackage{amsmath,amsfonts,amssymb}
\usepackage{calligra}
\usepackage{balance}
\usepackage{blkarray}

\newcommand{\V}[1] {\textsf{#1}} 			
\newcommand{\RV}[1] {\mathcal{#1}}			

\newcommand{\Bold}[1] {\bm{#1}} 			

\newcommand{\Sensor}[1]{
    \IfEqCase{#1}{
       {IW512}{L1}
       {IW576}{L2}
       {IW544}{L3}
       {IW554}{P1}
       {IW560}{P2}
        {#1}{#1}
    }
}

\newcommand{\TE}[1]{{\scriptsize\sffamily #1}} 
\newcolumntype{M}[1]{>{\centering\arraybackslash}m{#1}}
\newcolumntype{N}{@{}m{0pt}@{}}

\newcommand{\Eq}[1]{(\ref{#1})}

\newcommand{\Fig}[1]{Fig. \ref{#1}}
\newcommand{\Table}[1]{Table \ref{#1}}
\newcommand{\Sec}[1]{Section \ref{#1}}